\documentclass[conference]{IEEEtran}
\IEEEoverridecommandlockouts
\usepackage{cite}
\usepackage{amsmath,amssymb,amsfonts, amsthm}
\usepackage{graphicx}
\usepackage{textcomp}
\usepackage{xcolor}
\usepackage{booktabs}
\usepackage{threeparttable}

\usepackage[ruled,linesnumbered]{algorithm2e}
\usepackage{setspace}
\usepackage{makecell}
\usepackage{multirow}
\usepackage{colortbl} 
\usepackage{enumitem}

\theoremstyle{definition}

\newtheorem{exmp}{Example}

\def\BibTeX{{\rm B\kern-.05em{\sc i\kern-.025em b}\kern-.08em
    T\kern-.1667em\lower.7ex\hbox{E}\kern-.125emX}}
\begin{document}

\title{CounterCLR: Counterfactual Contrastive Learning with Non-random Missing Data in Recommendation%\author{\IEEEauthorblockN{Anonymous Author(s){*}}}}
% {\footnotesize \textsuperscript{*} Wenjia Wang is the corresponding author}
%}
\thanks{ This work was done when Jun Wang (jwangfx@connect.ust.hk) was a research intern at Kuaishou Technology Co., Ltd. Wenjia Wang (wenjiawang@ust.hk) is the corresponding author.}
}

\author{\IEEEauthorblockN{Jun Wang\textsuperscript{1}, Haoxuan Li\textsuperscript{2}, Chi Zhang\textsuperscript{3}, Dongxu Liang\textsuperscript{3}, Enyun Yu\textsuperscript{3}, Wenwu Ou\textsuperscript{3}, Wenjia Wang\textsuperscript{1, 4} }
\IEEEauthorblockA{$^1$\textit{Hong Kong University of Science and Technology}\\ 
$^2$\textit{Peking University}\\ 
$^3$\textit{Kuaishou Technology Co., Ltd.}\\
$^4$\textit{Hong Kong University of Science and Technology (Guangzhou)} 
% \\
}}
% \and
% \and
% \and
% \and
% \and

\maketitle
\IEEEpeerreviewmaketitle

\begin{abstract}
Recommender systems are designed to learn user preferences from observed feedback and comprise many fundamental tasks, such as rating prediction and post-click conversion rate (pCVR) prediction. However, the observed feedback usually suffer from two issues: \textit{selection bias} and  \textit{data sparsity}, 
{where biased and insufficient feedback seriously degrade the performance of recommender systems in terms of accuracy and {ranking}.}
Existing solutions for handling the issues, such as data imputation and inverse propensity score, are highly susceptible to additional trained imputation or propensity models.
In this work, we propose a novel \textbf{counter}factual \textbf{c}ontrastive \textbf{l}earning framework for \textbf{r}ecommendation, named CounterCLR, to tackle the problem of non-random missing data by exploiting the advances in contrast learning.
Specifically, the proposed CounterCLR employs a deep representation network, called CauNet, to infer non-random missing data in recommendations and perform user preference modeling by further introducing a self-supervised contrastive learning task. Our CounterCLR mitigates the selection bias problem without the need for additional models or estimators, while also enhancing the generalization ability in cases of sparse data. Experiments on real-world datasets demonstrate the effectiveness and superiority of our method.
\end{abstract}

\begin{IEEEkeywords}
recommendation system, non-random missing data, causal inference, contrastive learning.
\end{IEEEkeywords}

\section{Introduction}
\label{sec:intro}
In real-world recommender systems, users' interactive feedback on items such as rating and purchase is used to represent and develop user preferences. By using the observed sparse feedback matrix to inference the potential preferences or relevance of non-interactive user-item pairs, recommendation systems can be facilitated to provide diverse and personalized recommendations to users. Thus, the task of rating or conversion rate prediction has become the core problem and has attracted increasingly attention~\cite{wang2019doubly, wang2022escl, Chen-etal2021, liu2022kdcrec, saito2019towards, li2023res}.

% Hence, developing effective algorithms for rating prediction task is an important topic in both academia and industry.

Although many techniques have been proposed to tackle the missing data problem in recommendation, recent studies showed that the missing data prediction task usually suffers from two major issues: \textit{selection bias} and \textit{data sparsity}~\cite{ESMM, Chen-etal2022,  Wu-etal2022, wang2022escl}.
The selection bias, due to the users' self-selection behaviors or the policies of the recommender systems, results that the observed ratings are highly biased, which is widely known as data $\textit{Missing Not At Random}$ (MNAR) \cite{rubin1976inference,Wu-etal2022, gao2022causal, Zhang-etal-2022}. Specifically, since users tend to choose the items they like to rate or the recommender systems are more likely to recommend popular items to users, the higher (or the more positive) ratings are more likely to be observed. %Therefore, the observed MNAR data is significantly different with
% and there are systematic difference between data distribution in the observed ratings and 
%our target $\textit{Missing At Random}$ (MAR) population.
The data sparsity means that only a small portion of ratings are observed while the most of the ratings are missing.
It occurs since the interactions between the users and items are rare, compared with the number of entire user-item pairs. 
% For example, there are billions of video platforms like TikTok and Youtube, and tens of millions of videos are uploaded per day, while each user only spends less than two hours per day on the platform so that the user can only watch a small portion of the newly uploaded videos. 
% A similar occasion occurs in online-shopping platforms, where hundreds of millions of products are on the shelf while most users only have less than ten reviews.
% %Another example is that there are tens of millions of users and millions of items on a shopping platform, while most users only have less than ten reviews. 
% Two widely used public datasets in recommender systems literature, {Coat} and {Yahoo}, only have observed rates of $8\%$ and  $2\%$, respectively.
% %In reality, users' preferences to items are recorded as binary or multi-scaled ratings.
% However, recent studies shows that the ratings given by the users seriously suffer from the selection bias issue \cite{Chen-etal2022}, since 
% {\color{blue}The selection bias issue results that only the ratings to a small portion items are obsevered and the ratings to most other items are missing}, which is widely known as $\textit{Missing Not At Random}$ (MNAR) problem \cite{rubin1976inference}. 
In summary, the selection bias and data sparsity issues severely prevent recommender systems from learning users' true preferences and thus degrade the performance of the recommender systems.
% degrade the recommendation accuracy.

%In order to address the selection bias and data sparsity issues, many efforts are devoted to (). 
There are growing literature focusing on addressing the selection bias and data sparsity issues \cite{Chen-etal2022, Wu-etal2022}.
Among them, causality-based methods have become increasingly popular in recommender systems \cite{schnabel2016recommendations, wang2019doubly, saito2019doubly, guo2021enhanced, Dai-etal2022, li2023stabledr, li2023tdr, li2023propensity}, in which the missingness of feedback are modeled under the widely-used ``\textit{Potential Outcome Framework}'' (POF) in causal inference literature.
% Specifically, the observed ratings are the outcomes of the user-item pairs \textit{with} exposure treatments, while the rating prediction task is to answer the non-exposure question ``what the rating will be if exposing (recommending) an item to a user'' for the user-item pairs \textit{without} exposure treatments.
% The most promising 
One class of causality-based methods is reweighting-based methods, including inverse propensity score (IPS) \cite{schnabel2016recommendations, li2023propensity}, self-normalized inverse propensity score (SNIPS) \cite{schnabel2016recommendations}, and \text{doubly robust} (DR) estimation \cite{wang2019doubly,saito2019doubly,guo2021enhanced, Dai-etal2022, li2023stabledr, li2023tdr, li2023balancing, li2023propensity}.
Despite showing superior performance on debiasing, they highly rely on the accuracy of the propensity estimation, in which an extra model and/or unbiased datasets are required, and the performance degenerates with mis-specified model or sparse interacted data. 
%requires additional models to estimate the propensity, and the imperceptible perturbations from propensity estimations can handicap the propensity-based methods.
% {\color{blue} In particular, some models require an additional unbiased dataset for rating prediction...something like this?}

{
% As for data sparsity, most existing methods could address this challenge to a certain extent, and classical solutions includes collaborative filtering methods like matrix factorization (MF) \cite{koren2008factorization}, neural collaborative filtering (NCF), and its variants \cite{he2017neural, he2018outerncf, saitotowards, guo2013integrating}.
% % Typically, they elicit more pseudo user ratings by merging the ratings of reliable similar users, so as to provide sufficient data for training recommender systems.
% % However, these methods only exploit the observed ratings and overlook the difference between the distribution of the observed ratings and that of the entire ratings, which often leads to a biased estimation in consequence. 

As for data sparsity, many studies have proposed data augmentation-based methods, by eliciting more pseudo user ratings or merging the ratings of reliable similar users, so as to provide sufficient data for training the prediction model \cite{liu2022rating}.
However, these methods only exploit the observed feedback for data augmentation and ignore the distributional difference between the observed and potential feedback on all user-item pairs, which leads to biased predictions. 
Recent studies have proposed the use of multi-task learning \cite{ma2018entire, wang2022escm2, li2023res} to address both selection bias and data sparsity simultaneously. 
Different from these studies, we propose a novel contrastive learning approach to mitigate the data sparsity problem while addressing non-random missing data.

%Furthermore, most propensity-based methods only exploit the observed ratings for training and explicitly ignore the existence of data sparsity issue, which hinder their generalization performance for predicting ratings for the numerous unobserved user-item pairs.

%{\color{red} simply link with dragonnet}

%{\color{red} introduce how the data sparsity is sovled}

%{\color{red} introduce contrastive learning}

% Recently, contrastive learning as a self-supervised approach has drawn much attention for its superior performance on generalization ability and mitigating data sparsity issue \cite{yu2022self}, and show great advances in computer vision (CV)  and natural language processing (NLP) 
% \cite{chen2020simple, kong2019mutual, liu2022selfsupervised}
% %Unsurprisingly, this method has also been applied to recommender systems, and improved the performance of recommender systems in multiple aspects \cite{zhang2021causerec, yao2021self, wu2021sigir}, {such as alleviating popularity bias and learning more robust user representation.}
% Nevertheless, to the best of our knowledge,
% % as far as we know, 
% there is a lack of works that apply self-supervised contrastive learning to mitigate the selection bias and data sparsity issues in rating prediction task.

In this work, we
% Furthermore, the two key differences motivate us to 
propose a \textbf{Counter}factual \textbf{C}ontrastive \textbf{L}earning framework for \textbf{R}ecommendations, named \textbf{CounterCLR}, to address the non-random missing data and data sparsity issues simultaneously.  
The proposed CounterCLR is composed of a causality-based network named {CauNet} and a contrastive learning auxiliary task. 
{
% We analysis the generation process of selection bias and data sparsity in rating prediction task under the Potential Outcome Framework and 
Specifically, the {CauNet} is designed under the potential outcome framework to model the non-random missing feedback from the observed interactions.
The contrastive learning objective in CauNet conducts an auxiliary task for learning user and item representations, which enhances the generalization ability in cases of sparse data.

% By using the cascade structure between CauNet and contrastive learning objective, the CounterCLR can fully leverage multiple feedbacks, including observed ratings and predicted ratings from CauNet.

% Specifically, the proposed CounterCLR employs a deep representation network, called CauNet, to infer non-random missing data in recommendations and perform user preference modeling by further introducing a self-supervised contrastive learning task. Our CounterCLR mitigates the selection bias problem without the need for additional models or estimators, while also enhancing the generalization ability in cases of sparse data. Experiments on real-world datasets demonstrate the effectiveness and superiority of our method.
% Furthermore, the contrastive learning objective is in a cascade of the CauNet, in order to fully use the multiple feedbacks, including observed ratings and predicted ratings from CauNet.
% It conducts an auxiliary task for learning user preference representation. 
% and is able to reduce the selection bias. 
% {\color{blue}Furthermore, in order to make better use of the ratings, the contrastive learning objective conducts an auxiliary task on learning user preference representation.
% It can not only improves the generalization ability of {CauNet} under data sparsity, but also further alleviates the selection bias.} 
The contributions of this paper are summarized as follows.
\textbf{i)} It not only exhibits a substantial enhancement of generalization ability under sparse interactions, but also alleviates the problem of non-random missing data, thus can address selection bias and data sparsity issues simultaneously in recommender systems. 
\textbf{ii)} 
Our framework does not require any additional models or unbiased datasets, and thus it is practically preferred. 
% Furthermore, it 
% Moreover, it has noticeable advantages in rating prediction, and 
% reduces the reliance on the quality of propensity estimation.
% is less sensitive to the accuracy of 
% Moreover, our numerical experiments suggest that CounterCLR has ;
% \textbf{iii)} 
{\textbf{iii)} }
A novel casually contrastive paradigm is proposed for debiased recommendation and user preference modeling.
%Finally, the effectiveness of {CounterCLR} is fully demonstrated  empirically. 
Extensive experiments on real-world datasets illustrate the above merits of CounterCLR, and show that {CounterCLR} substantially outperforms state-of-the-art methods.} 
% A rigorous proof is further provided explaining how the contrastive learning objective works in essence, consolidating the efficiency of {CounterCLR} from the theoretical perspective. 

\section{Related Work}
\label{sec:relate work}
In this section, we review many previous methods designed for debiased recommendation and recent advances on self-supervised contrastive learning (SSCL).
% for not only recommendation but other domains as well.

%这里要提别的rating methods吗？

% \paragraph{Debiased Recommendation}
% \subsection{Debiased Recommendation}
% \label{subsec: debiased rec}
\textbf{Debiased Recommendation.}
% Since the rating prediction is a crucial task of learning user preference in recommender system, a substantial amount studies have been developed for debiasd rating prediction.
% In the earlier study, researchers focus on 
The inconsistency between training and test set distributions has been widely studied~\cite{wang2022invariant, liu2022kdcrec, du2022invariant, huang2023pareto, wang2023ot, wang2023out, wang2022estimating, wang2023treatment, zou2023factual}.
Existing methods tackling the non-random missing data can be divided into three categories: generative modeling-based, relabeling-based, and reweighting-based methods. 
By assuming a data generation process and adjusting for biases accordingly, generative modeling-based methods \cite{hernandez2014probabilistic, wang2023generative} leverage heuristic human prior knowledge and provide an explainable solution for debiasing. 
% To address the non-random missing data, previous methods exploited the heuristic that a rating's latent value affects its probability of being missing, introducing an additional matrix factorization model to capture this biased missingness, and assist for debiased prediction via a joint-training manner. However, training generative models can be challenging and requires strong assumptions about the underlying data generation process.
Relabeling methods \cite{steck2010training, wang2019doubly}  typically mitigate the selection bias by data imputation, which adaptively downweights the contribution of imputed ratings for unobserved user-item pairs to the loss function.
% previous methods are developed based on the fact that users are more likely to supply ratings to their favorite items.
% However, these methods can be inefficient and highly sensitive to the quality of the pseudo-ratings. 
Reweighting methods involve assigning weights induced from the propensity scores to each instance, in order to rescale their contributions during model training.
Typical reweighting methods include IPS \cite{schnabel2016recommendations, li2023propensity}, self-normalized inverse propensity score (SNIPS) \cite{schnabel2016recommendations}, and DR \cite{wang2019doubly,saito2019doubly,guo2021enhanced, Chen-etal2021, Dai-etal2022, li-etal-MR2023, Li-etal2023KDD, song2023cdr},
% the inverse propensity scoring (IPS) and self-normalized inverse propensity scoring (SNIPS) methods \cite{schnabel2016recommendations}, 
where the propensity scores are obtained through naive Bayes or logistic regression.
Nevertheless, the above methods highly depend on additional imputation models or propensity estimators, which need to be trained separately.

{On the other hand, there have also been several works that avoid the introduction of extra models or estimators. Specifically,
\cite{saito2020asymmetric} developed a model-agnostic meta learning method by deploying the asymmetric tri-training framework for unsupervised domain adaptation.
% This method first employed two pre-trained rating models to generate a less-biased pseudo rating set, and then trained the target rating model on this pseudo set to obtain debiased recommendation.
\cite{wang2020information, liu2021mitigating} built information-theoretic frameworks, where novel non-exposure variational information bottlenecks are derived for addressing the non-random missing data problem.}
Alternatively, \cite{liu2022rating} proposed a self-supervised learning-based method, in which an extra collected unbiased \textit{Missing At Random} (MAR) rating data is required to calibrate the rating distribution.}
However, the collection of unbiased MAR ratings can be costly or even unavailable in practice. %(Any other works?) 
In contrast, in this work, 
% with the aforementioned works, 
we proposed a causally contrastive learning-based framework that essentially introduces neither additional models and estimators nor extra unbiased MAR data, which is more practically preferred.

%requires no introduction of neither additional models and estimators nor extra unbiased MAR data, is more practically preferred. 
% Specially, we integrate a casual rating model with a self-supervised contrastive learning objective,  which is trained on only the observed data,  so as to reduce the affect of selection bias.
% Hence, our method are more practical and robust.
% than existing methods.

% \paragraph{Self-supervised Contrastive Learning}
\textbf{Self-supervised Contrastive Learning.}
% \subsection{Self-supervised Contrastive Learning}
% \label{subsec:scl}
% As a dominant component in self-supervised learning, 
Self-supervised contrastive learning (SSCL) aims to boost the model generalization ability by representation learning, where the embeddings of the augmented versions of the same sample are trained to be close to each other, while those of different samples are required to be pushed away.
This method has been widely utilized in CV and NLP areas \cite{chen2020simple, kong2019mutual, liu2022selfsupervised}. %and recently introduced in  recommender systems field \cite{zhang2021causerec, yao2021self, wu2021sigir} {to improve the recommendation performance in many aspects, including mitigating popularity bias and boosting robust user representation learning.}
%attracted interests of various domains such as CV and NLP\cite{he2017neural, chen2020simple} and nature language processing \cite{kong2019mutual, clark2020pre}.
%Recently , \cite{zhang2021causerec, yao2021self, wu2021sigir} apply this method in recommender systems field. 
% {Many studies \cite{liu2022selfsupervised, zhou2020cikm, zhou2021contrastive} also demonstrate the effectiveness of self-supervised contrastive learning in addressing domain adaption.
\cite{liu2022selfsupervised} empirically and theoretically shows that self-supervised contrastive learning can learn more representative features for helping classification task in long-tailed labeled image datasets. 
Recently, \cite{xie2022contrastive, zhou2020cikm, cai2023lightgcl} studied the contrastive learning methods in sequential recommendations to learn better item representations in the presence of long-tail items. \cite{zhou2021contrastive} exploited a SSCL-based method, named CLRec, to alleviate the popularity bias in deep candidate generation. However, these methods failed to tackle the problem of non-random missing data in recommendation.
% In addition,
%that self-supervised contrastive learning also show promising capability for debiasing learning.
%In the domain of computer version, \cite{liu2022selfsupervised} empirically and theoretically shows that self-supervised contrastive learning can learn more representative feature for promoting classification task in long-tailed labeled image dataset.
%\cite{zhou2021contrastive} exploited a SSCL based recommender named CLRec to alleviate the popularity bias in deep candidate generation. They also bridged the theoretical gap between contrastive learning objective and traditional unbiased recommendation objective, i.e., Info-NCE loss and inverse propensity weighted loss.}

\section{Preliminaries}
\label{sec: preliminary}
In this section, we formulate the problem of non-random missing data in recommendation using the widely-adopted $\textit{Potential Outcome Framework}$ (POF) in causal inference literature, and introduce the motivation of the proposed framework (CounterCLR) from a causal inference view. 
%Specifically, we reformulate the rating prediction task under the $\textit{Potential Outcome Framework}$ (POF), which is a classic analysis framework for causal inference problems.

% provide a causal inference view of recommendation and the motivation of the proposed framework CounterCLR by reformulating rating prediction task under the $\textit{Potential Outcome Framework}$ (POF), which is a classic analysis framework for causal inference problems.
% {\color{blue}Then, from the perspective of causal inference, we reformulate the rating prediction task and demonstrate our intuitions under the $\textit{Potential Outcome Framework}$ (POF) }.

\subsection{Problem Setup}
% \label{subsec:formulation}
%基本nation + 基本模型的介绍（MF/NCF）

% Firstly, we introduction some classical concepts for modelling rating prediction task in recommendation systems.
Let the user set and item set be $\mathcal{U} = \{u_{1}, u_{2}, \cdots, u_{N}\}$ and $\mathcal{I} = \{i_{1}, i_{2}, \cdots, i_{M}\}$, respectively, and the set of all user-item pairs be $\mathcal{D} = \mathcal{U} \times \mathcal{I}$.
% Let $\mathcal{E}\subset \mathcal{D}$ such that $(u,i)\in \mathcal{E}$ denotes the item $i$ is exposed to the user $u$.
Let $\mathbf{R} \in \mathbb{R}^{N \times M}$ be the matrix of true feedback (e.g., rating, conversion, etc.), with elements $r_{u, i}$ be the true feedback of user $u$ to item $i$.
For modeling the missing mechanism of the observed feedback, 
we introduce an observing indicator matrix 
% let 
$\mathbf{O} = (o_{u,i}) \in \mathbb{R}^{N \times M}$, where $o_{u, i} = 1$ represents that $r_{u, i}$ is observed or missing $o_{u, i} = 0$.
% It should be noted that if $r_{u, i}$ is observed, then the item $i$ must be exposed to the user $u$, i.e., $(u,i)\in \mathcal{E}$. If $r_{u, i}$ is not observed, then $(u,i)$ may or may not be in $\mathcal{E}$, because it is possible that the item $i$ has been exposed to the user $u$, while the user $u$ does not rate it. 
%In the rating prediction task, 
% In the rating prediction task, the problem of interest is "what would the user's rating $r_{u, i}$ be if the item $i$ had recommended to the user $u$". However, due to the non-random recommendation exposure and user selection, the collected data contains only sparse explicit ratings. In the post-click conversion rate (pCVR) prediction task, the problem of interest is "what would the user's conversion indicator $r_{u, i}$ be if user $u$ had clicked on the item $i$". However, users selectively click on the items, and the collected data only contains $r_{u, i}$ in the occurrence of the click behavior $o_{u, i}=1$.
% To tackle the above problem, 
For the rating or the post-click conversion rate prediction task,
we aim to train a prediction model %$\mathcal{M}$
% $\mathcal{M}: (u, i) \to \hat{r}_{u, i}$
minimizing the training loss
\begin{align}
    {\rm Error}(\mathbf{\hat{R}}, \mathbf{R}) = \sum_{(u,i) \in \mathcal{D}}\ell(\hat{r}_{u, i}, r_{u, i}),
\end{align}
where $\hat{\mathbf{R}} = (\hat{r}_{u, i}) \in \mathbb{R}^{N \times M}$ is the predicted rating matrix, and $\ell(\cdot,\cdot)$ is a loss function, e.g., squared loss. 

% \alert{In order to} 

In practice, recommender systems conceptually model the user $u$ and item $i$ by $K$-dimensional embeddings, i.e.,
% $\mathbf{e}_{u} \in \mathbb{R}^{K}$ and $\mathbf{e}_{i} \in \mathbb{R}^{K}$
$\mathbf{e}_{u},\mathbf{e}_{i}\in \mathbb{R}^{K}$, respectively. %\footnote{The dimensions of $\mathbf{e}_{u}$ and $\mathbf{e}_{i}$ are the same here for simplicity.}. 
% The embeddings $\mathbf{e}_{u}$ and $\mathbf{e}_{i}$ represent the user features and item features, respectively.
% \alert{These embeddings can be learned via...}
Then, each user-item pair $(u, i)$ is represented by the embedding concatenation as $\mathbf{x}_{u,i} = (\mathbf{e}_{u}, \mathbf{e}_{i})$.
Two common rating prediction models in recommender systems are matrix factorization (MF) \cite{koren2009matrix} and neural collaborative filtering (NCF) \cite{he2017neural}.
The MF model directly processes $\mathbf{x}_{u, i}$ to conduct the rating prediction by $\hat{r}_{u, i} = \mathbf{e}_{u}^{\top}\mathbf{e}_{i}$, while the NCF model applies a multi-layer feedforward neural network to obtain the feedback prediction $\hat{r}_{u, i}$.

% sequentially processes $\mathbf{x}_{u, i}$ through multiple hidden layers to . \alert{NCF neural networks?}

% \subsection{A Causal Inference Perspective}
\subsection{Problem Formulation under POF}
% \label{subsec: causalperspective}
%POF reformulate RPT 并讲述我们的intuition

%1. only R(1) is observed
%2. for a individual unit, R(1) R(0) are similar since they are determined by the user preference; however, from population level, there is a covariate shift between R(1) and R(0)

% The last section first reformulates the rating prediction task under POF, and then provides a discussion on two intrinsic disconnects of the rating prediction task to the standard problems in POF, so as to demonstrate our intuitions.

% \alert{need to introduce POF here}
Conceptually, the POF is formulated by three components: the covariate set $\mathcal{X}$, the treatment set $\mathcal{T}$, and the potential outcome set $\mathcal{Y}$.
For example, for a diabetic patient with covariate $x \in \mathcal{X}$, doctors are usually interested in whether a new medical treatment affects the blood sugar level $y \in \mathcal{Y}$. The potential outcomes in this example can be modeled as $y(0), y(1)\in \mathcal{Y}$, and $\mathcal{T}=\{0, 1\}$ with $0$ and $1$ denote the patient not take and take this new medication, respectively.
% an assignment of a new pose, i.e, $\mathcal{T} = \{0, 1\}$, whether affects 

Similarly, the missing mechanism in recommendation can be modeled by an treatment assignment mechanism in the POF. To see this, note that $o_{u, i} = 1$ can be treated as assigning an exposure treatment to the user-item pair $(u, i)$, i.e., the item $i$ is recommended to the user $u$.
% \footnote{We assume that as long as an item $i$ is recommended to a user $u$, then $o_{u, i} = 1$ and $r_{u,i}$ is observed.}. 
Consequently, we can follow the POF to redefine the rating prediction task as follows.

% in the following Definition~\ref{def:rpPOF}. 
\begin{exmp}[Rating Prediction Task under POF]
% \begin{defn}[Rating Prediction Task under POF]
\label{def:rpPOF}
For each user-item pair $(u, i)$, $o_{u, i}$ is the treatment indicator and there are two potential outcomes $r_{u, i}(1)$ and $r_{u, i}(0)$, named \textit{exposure rating} and \textit{non-exposure rating}, respectively.
% (denoted by $r_{u, i}^{1}$ and $r_{u, i}^{0}$ for simplicity). 
The rating prediction task is to estimate the {exposure} ratings of users to all items, i.e., estimate $r_{u, i}(1)$ for all $(u, i) \in \mathcal{D}$.
\end{exmp}
% \end{defn}

For simplicity, hereafter we denote $r_{u, i}(1)$ and $r_{u, i}(0)$ as $r_{u, i}^{1}$ and $r_{u, i}^{0}$, respectively.
In the above example, the observed ratings are {the exposure ratings}, i.e., the potential outcomes of the units with the exposure treatment. Meanwhile, the ratings of the units without the exposure treatment are missing.
% The true rating set can be considered as $\mathcal{R} = \{ (\mathbf{x}_{u, i}, o_{u, i}, r_{u, i}): (u, i) \in \mathcal{D} \}$, and the observed rating set and the missing rating set can be written as $\mathcal{R}^{1} = \{ (\mathbf{x}_{u, i}, o_{u, i}, r_{u, i}): o_{u, i} = 1, (u, i) \in \mathcal{D}\}$  and $\mathcal{R}^{0} = \{ (\mathbf{x}_{u, i}, o_{u, i}, r_{u, i}): o_{u, i} = 0, (u, i) \in \mathcal{D} \}$, respectively.
% The true rating set can be considered as $\mathcal{R} = \{ (\mathbf{x}_{u, i}, o_{u, i}, r_{u, i}): (u, i) \in \mathcal{D} \}$, and the observed rating set and the missing rating set can be written as $\mathcal{R}^{1} = \{ (\mathbf{x}_{u, i}, o_{u, i}, r_{u, i}): o_{u, i} = 1, (u, i) \in \mathcal{D}, \}$  and $\mathcal{R}^{0} = \{ (\mathbf{x}_{u, i}, o_{u, i}, r_{u, i}): o_{u, i} = 0, (u, i) \in \mathcal{D} \}$, respectively.
% Namely, in rating prediction task, only the potential outcomes of the units with action 1 (i.e., exposure treatment) are observed.
% Consequently, the rating prediction task under the umbrella of POF is to infer the missing data $\mathcal{R}^{0}$ only using observed data $\mathcal{R}^{1}$.
% Based on Definition~\ref{def:rpPOF},
%The causal graph of rating prediction task is illustrated in Figure~\ref{fig:causalgraph}. 
This motivates us to introduce a causal rating prediction model CauNet into the rating prediction task.
Notably, the rating prediction task under POF is different with the traditional POF problems like investigating the therapeutic effects of different treatments as in the previous diabetes example, and two challenges are raised from the differences in consequence. 
% We then demonstrate how the selection bias and data sparsity occur from a causal inference aspect.
% We argue that there are two important distinctions between rating prediction task and standard POF problems like investigating the therapeutic effects of different treatments in medical study. 
First, since users tend to rate the items they like, the higher ratings become more likely to be observed. Therefore, the collected user-item pairs with observed ratings could not be used as a representative of the target population, i.e., the entirety of the user-item pairs, which leads to the sample selection bias.
% Therefore, the assignment mechanism of exposure treatment in the rating prediction task depends on the potential outcomes, i.e, there should be a causality link from $r_{u, i}$ to $o_{u, i}$, which leads to the selection bias.
% Specially, since users tends to choose the items they like to rate, the higher ratings become more likely to be observed, which cause the selection bias issue.
Second, only the potential outcomes $r_{u, i}^{1}$ are observable, while $r_{u, i}^{0}$ are unavailable if the item $i$ is not recommended to the user $u$, which arises the data sparsity issue. These challenges call for new methods for modeling the causalities in the rating prediction task.

% bring two challenges, selection bias and data sparsity, to the rating prediction task. 

% To circumvent the two critical challenges, we design our CounterCLR method 

% \begin{figure}[h]
% \center
% \vspace*{-2mm}
%   \includegraphics[width=0.25\linewidth, height = 0.2\linewidth]{pictures/causalgraph.pdf}
%   \vspace*{-1mm}
%   \caption{The causal graph of non-random missing data in recommendation.}
%  \label{fig:causalgraph}
% %  \vspace*{-3mm}
% \end{figure}

\section{Methodology}
\label{sec:method}

\begin{figure*}[h]
\center
\includegraphics[width=1\linewidth, height = 0.27\linewidth]{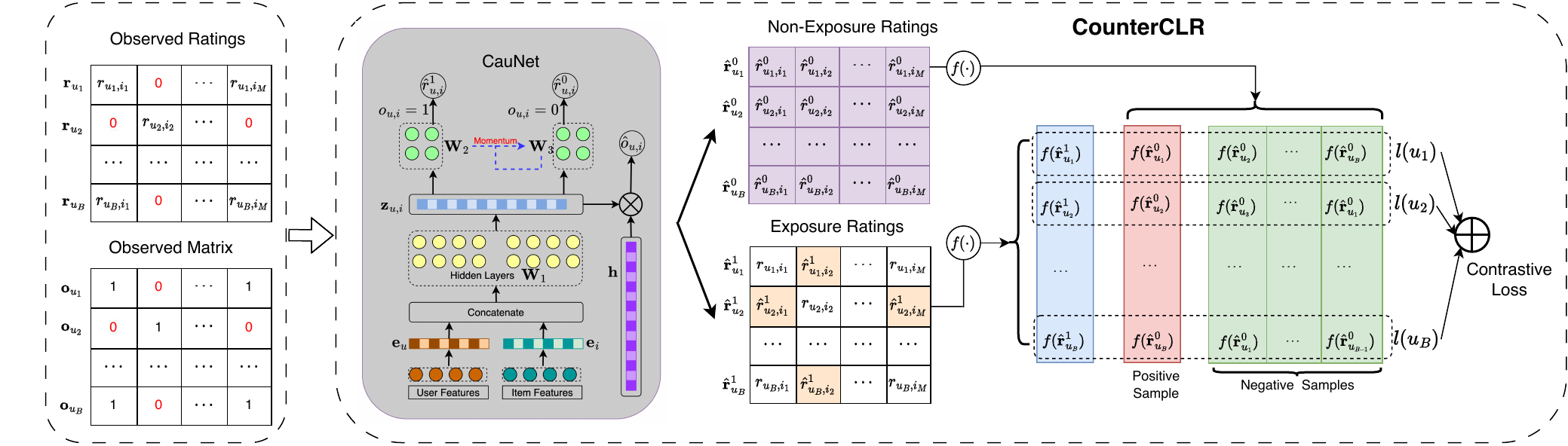}
%   \vspace*{-1mm}
  \caption{The architecture of CounterCLR, including a causality-based prediction model with an auxiliary contrastive learning objective.}
 \label{fig:conloss}
\vspace*{-3mm}
\end{figure*}

%
%%connections between CauNet & CLO
%1. momentum 与CLO
%2. r^{1} r^{0} 与 CLO

In this section, we introduce the proposed causality-based contrastive learning framework for addressing non-random missing data in recommendation, named CounterCLR, which also greatly alleviates the data sparsity issue simultaneously. The CounterCLR consists of two parts: (1) a causality-based prediction model and (2) a contrastive learning objective. 

\subsection{Causality-Based Prediction}
\label{subsec: causalpred}
%去学习propensity的task 本质上是在balance R(1)与R(0)的covariate
% {\color{green}In order to model the causalities of rating prediction task shown in Figure~\ref{fig:causalgraph}, inspired by Dragonnet \cite{shi2019adapting}, we propose a three-headed architecture CauNet, which not only predicts the potential ratings but also estimates the propensity score. It should be noted that the estimated propensity score is not passed to the contrastive learning objective, thus weakly influences the rating prediction performance. Therefore, as we will see later, a simple propensity score estimation procedure is sufficient, which reflects a nice property of CounterCLR: weak dependency on the accuracy of propensity score estimation. This estimation, albeit not necessarily \textit{very accurate}, can help to learn a better representation and reduce the selection bias, thus improve the performance of CounterCLR.}

{
% In order to model the non-random missing data from the observed interactions, we adopt a three-headed architecture, namely CauNet. 
The CauNet is a three-headed architecture, which can predict the exposure and non-exposure ratings, i.e., predict ${r}_{u, i}^{1}$ and ${r}_{u, i}^{0}$, and also estimate the propensity score with a simple and efficient procedure at the same time. 
The predicted exposure and non-exposure ratings are passed to the contrastive learning objective to construct positive and negative pairs. % (to be introduced in Section \ref{subsec: clo}).
In addition, the propensity score estimation can help to learn a better representation for predicting the treatment and reduce the selection bias.}

The processing steps in CauNet are presented in Figure~\ref{fig:conloss}.
Given an embedding concatenation $\mathbf{x}_{u,i} = (\mathbf{e}_{u}, \mathbf{e}_{i})$, we first use a neural network for encoding $\mathbf{x}_{u, i}$ to $\mathbf{z}_{u, i}\in \mathbb{R}^p$, i.e., $\mathbf{z}_{u, i} = h_{\mathbf{W}_{1}}(\mathbf{x}_{u, i})$. With the feature embedding $\mathbf{z}_{u, i}$, we use another two neural networks to {predict the exposure and non-exposure ratings} $\hat{r}_{u, i}^{1}$ and $\hat{r}_{u, i}^{0}$ {when $o_{u, i} = 1$ and $o_{u, i} = 0$, respectively, i.e., 
$
    \hat{r}_{u, i}^{1} = h_{\mathbf{W}_2}(\mathbf{z}_{u, i},1)$ and
    $\hat{r}_{u, i}^{0} =
    h_{\mathbf{W}_3}(\mathbf{z}_{u, i},0).
$
% , where $\mathbf{W}_{1}$ is the parameter, and the dimension of $\mathbf{W}_{1}$ depends on the architecture of the neural network.
% i.e, $\mathbf{z}_{u, i} = \text{ReLU}(\mathbf{W}_{1}^{\top}\mathbf{x}_{u, i})$. With the feature embedding $\mathbf{z}_{u, i}$, we use two 1-hidden layer networks to predict the corresponding potential rating $\hat{r}_{u, i}^{1}$ or $\hat{r}_{u, i}^{0}$ when $o_{u, i} = 1$ or $o_{u, i} = 0$, respectively, i.e., 
% \begin{equation*}
%     \hat{r}_{u, i}^{1} = \text{ReLU}(\mathbf{W}^{\top}_{2}\mathbf{z}_{u, i}), \quad
%     \hat{r}_{u, i}^{0} =
%     \text{ReLU}(\mathbf{W}^{\top}_{2}\mathbf{z}_{u, i}).
% \end{equation*}
Here, $\mathbf{W}_{1}, \mathbf{W}_{2}$, and $\mathbf{W}_{3}$ are the parameters inside the neural networks $h_{\mathbf{W}_1}$,  $h_{\mathbf{W}_2}$, and $h_{\mathbf{W}_3}$, respectively.
% and the dimensions of $\mathbf{W}_{1}, \mathbf{W}_{2}$ depend on the architecture of the corresponding neural networks. 
% Note that the two neural networks for rating prediction share the same parameter $\mathbf{W}_{2}$, of which the reason will be explained later.
Since the true ratings are determined by the user preference and should not be influenced by the exposure treatment in the rating prediction task, 
% the distributions of $r_{u, i}^{1}$ and $r_{u, i}^{0}$ should be similar for a user-item pair $(u, i)$. Therefore,
we update $\mathbf{W}_3$ in a momentum manner, i.e., $\mathbf{W}_3 \leftarrow m \mathbf{W}_{3} + (1 - m) \mathbf{W}_{2}$, to ensure that $h_{\mathbf{W}_2}$ and $h_{\mathbf{W}_3}$ are close, where $m$ is a hyper-parameter controlling the weights of $\mathbf{W}_{2}$ in the momentum update.
% Note that $m=0$ implies that  $h_{\mathbf{W}_3}$ shares the parameters with $h_{\mathbf{W}_2}$. 
Additionally, in Section~\ref{subsec: clo}, we will see this momentum update mechanism cooperates with the contrastive learning objective to help minimize the discrepancy between the distributions of $r_{u, i}^{1}$ and $r_{u, i}^{0}$.
}

{Since the propensity score estimates are not passed to the contrastive learning objective and do not explicitly influence the quality of the CounterCLR, we can construct a rather simple model for the propensity score estimation. 
% For propensity score estimation, 
Motivated by the random feature regression models \cite{yang2021exact,mei2022generalization}, we use a linear transformation with a random vector $\mathbf{h} \in \mathbb{R}^p$ followed by a sigmoid function, i.e., $\hat{o}_{u, i} = \operatorname{Sigmoid}(\mathbf{h}^{\top}\mathbf{z}_{u, i})$. % and $\operatorname{Sigmoid}(x) = 1/(1 + \exp(-x))$. 
Note that $\mathbf{h}$ is fixed after initialization.
In summary, the model parameter of CauNet is $\theta = (\mathbf{W}_{1}, \mathbf{W}_{2}, \mathbf{W}_{3})$, which is trained to minimize
%\begin{equation}\label{eq: cauloss}
\begin{align}
     \mathcal{L}_{cau}  &= 
    \mathcal{L}_{base} + \alpha \mathcal{L}_{pro}, \label{eq: cauloss} %\\ \nonumber
    %  \text{where~~} \mathcal{L}_{base} &= \sum_{(u,i) \in \mathcal{O}^{1}} (\hat{r}_{u, i}^{1} - r_{u, i})^{2}/\hat o_{u, i},\\ \nonumber
    % \mathcal{L}_{pro} &= \sum_{(u, i) \in \mathcal{D}}\text{CE}(\hat{o}_{u, i}, o_{u, i}). \nonumber
\end{align}
%\end{equation}
where $\mathcal{L}_{base} = \sum_{(u,i) \in \mathcal{O}^{1}} (\hat{r}_{u, i}^{1} - r_{u, i})^{2}/\hat o_{u, i}$, and $\mathcal{L}_{pro} = \sum_{(u, i) \in \mathcal{D}}\text{CrossEntropy}(\hat{o}_{u, i}, o_{u, i})$.
Hereby $\mathcal{O}^1 = \{(u,i)\in \mathcal{D}: o_{u,i} = 1\}$, and $\alpha$ is a hyper-parameter.

\subsection{Contrastive Learning Objective}
\label{subsec: clo}
% As discussed in the introduction, the ratings are the representations of the users' preference and the selection bias are caused by the
% ratings 是user preference的一种表征， selection bias 导致ratings有缺失， 使得用户表征的也是biased

%要不要把rating distribution feature 表述成 user preference representation?
%如果不： 理论解释那一部分可能不太好表述
%如果是： define loss这一部分就会有点绕( ratings 是一种user profile, 即可以视为user preference embedding/user embedding)
% As stated in Introduction, the ratings of the users represent the user preference. 
% {\color{blue} To address the data sparsity issue and further reduce the selection bias in the rating distribution and , we leverage SSCL on user preference embeddings extracted from the ratings.}

{In order to better learn the causality in rating prediction task, we deploy a contrastive learning objective cascading to CauNet.}
% Meanwhile, this objective can effectively harness the predicted exposure/non-exposure ratings in a self-supervised learning manner to alleviate the data sparsity issue.
% {\color{blue}In order to fully use the rating predictions $\hat{r}_{u, i}^{1}$ and $\hat{r}_{u, i}^{0}$, we deploy a contrastive learning objective cascading to CauNet.}
The contrastive learning objective
leverages SSCL on user preference embeddings extracted from the ratings
%, and it}  
% consists of two key steps.
% First, in order to build a contrastive learning objective, representative user preference embeddings are extracted from the ratings of users. 
% Then, we design a contrastive learning loss following the principle in SSCL, i.e., the user preference embeddings from the {exposure and non-exposure}  ratings of the same users (positive pairs) are close, while those of different users (negative pairs) are dissimilar.
% \paragraph{User Preference Extraction} 
First, let $\hat{\mathbf{r}}_{u} = [\hat{r}_{u,i_{1}},  \cdots, \hat{r}_{u, i_{M}}]$ stores the predicted ratings of a user $u$ to all items. 
An ideal user preference embedding should not only be informative to the ratings but also reserve the statistical properties of the rating distribution.
Hence,
%for comprehensively distilling the user preference embeddings from the predicted ratings $\hat{\mathbf{r}}_{u}$,
we adopt the aggregation function $f(\cdot)$ proposed by \cite{liu2022rating} as a user preference extractor, for mapping the predicted ratings $\hat{\mathbf{r}}_{u}$ to the user preference embedding $f_{\hat{\mathbf{r}}_{u}}$ as 
\begin{equation} \label{eq: feature}
\begin{aligned}
    f_{\hat{\mathbf{r}}_{u}} =   f(\hat{\mathbf{r}}_{u}) =& [f^{(1)}(\hat{\mathbf{r}}_{u}), \cdots, f^{(K)}(\hat{\mathbf{r}}_{u})], 
\end{aligned}
\end{equation}
where $K$ is the dimension of the user embedding $\mathbf{e}_{u}$, $ f^{(k)}(\hat{\mathbf{r}}_{u}) = \frac{1}{M} \sum_{i = 1}^{M}\sigma \left(\tau \left(\frac{k}{K}r_{\max} + \frac{K - k}{K} r_{\min} - \hat{r}_{u, i}\right)\right)$,
% \begin{align*}
%     f^{(k)}(\hat{\mathbf{r}}_{u}) = \frac{1}{M} \sum_{i = 1}^{M}\sigma \left(\tau \left(\frac{k}{K}r_{\max} + \frac{K - k}{K} r_{\min} - \hat{r}_{u, i}\right)\right),
% \end{align*}
$r_{\max}$ and $r_{\min}$ are the maximum and minimum possible ratings respectively, $\sigma(\cdot)$ is the sigmoid function, and $\tau$ is the scale parameter. It has been proved by \cite{liu2022rating} that the user preference extractor $f(\cdot)$ is indeed an approximation of the empirical cumulative distribution function of $\mathbf{\hat{r}}_{u}$.
{In addition, Eq.~(\ref{eq: feature}) is differentiable and adapted to gradient-based optimization methods, and theoretically guarantees the user preference embeddings and rating vectors share the same distributions between observed and unobserved user-item pairs. }
% \vspace{-0.2mm}

{Next, we define the exposure 
rating vector of the user $u$ as $\mathbf{\hat{r}}^{1}_{u} = [o_{u, i_{1}}r_{u, i_{1}} + (1 - o_{u,i_{1}})\hat{r}^{1}_{u,i_{1}},\cdots, o_{u, i_{M}}r_{u,i_{M}} + (1 - o_{u, i_{M}}) \hat{r}^{1}_{u, i_{M}}]$, 
and his/her non-exposure rating vector as
$\mathbf{\hat{r}}^{0}_{u} = [\hat{r}^{0}_{u,i_{1}},\cdots, \hat{r}^{0}_{u, i_{M}}]$.
Through the user preference extractor, we can obtain the exposure user preference embedding $f_{{\mathbf{\hat{r}}}^{1}_{u}} = f(\mathbf{\hat{r}}^{1}_{u})$ and the non-exposure one  $f_{{\mathbf{\hat{r}}}^{0}_{u}} = f(\mathbf{\hat{r}}^{0}_{u})$.}
{
Then, we define {our causally contrastive loss}  as
\begin{equation}\label{eq: conloss}
    \mathcal{L}_{con} = \sum_{u \in \mathcal{U}} \ell(u)  =
    \sum_{u \in \mathcal{U}} -\log\left(\frac{\exp(f_{{\hat{\mathbf{r}}}_{u}^{1}}^{\top}f^{}_{\hat{\mathbf{r}}_{u}^{0}}/t)}{\sum_{u' \in \mathcal{U}}\exp(f_{\hat{\mathbf{r}}_{u}^{1}}^{\top}f^{}_{\hat{\mathbf{r}}_{u'}^{0}}/t)}\right),
\end{equation}
% potential ratings without exposing; potential ratings with exposing and observed ratings
where $t$ is called temperature hyper-parameter.
% \vspace{-0.5mm}

An illustration of the contrastive learning objective is presented in Figure~\ref{fig:conloss}.
In practice, $\mathcal{L}_{con}$ is optimized in a batch-wise manner.
Notice that, as previously highlighted in Section \ref{subsec: causalpred}, the distributions of $\hat{r}_{u, i}^{1}$ and $\hat{r}_{u, i}^{0}$ should be close,
{indicating that the exposure and non-exposure user preference embeddings, i.e. $f_{\hat{\mathbf{r}}_{u}^{1}}$ and $f_{\hat{\mathbf{r}}_{u}^{0}}$, should be similar.}
% for the same user.} 
% Hence, in $\mathcal{L}_{con}$, we treat the exposure and non-exposure user preference embeddings  of the same user as positive pairs and those of different as negative pairs. 
% More specifically, b
To this end, by minimizing $\mathcal{L}_{con}$, we can pull closer $f_{\hat{\mathbf{r}}_{u}^{1}}$ and $f_{\hat{\mathbf{r}}_{u}^{0}}$
% the exposure and non-exposure user preference embeddings of the same user 
for the user $u$,
and push away
$f_{\hat{\mathbf{r}}_{u}^{1}}$ and $f_{\hat{\mathbf{r}}_{u'}^{0}}$ 
% those of different users 
for $u'\neq u$.

\begin{table*}[h]\footnotesize
% \small
\centering

\setlength{\tabcolsep}{6pt}
%\begin{spacing}{1}
\caption{The MSE, MAE and nDCG@5 on three real-world datasets. The best results among the methods without unbiased MAR data are bold. The best results among all methods are underlined.  NRU represents ``not require unbiased MAR data''.}
\vspace{-2pt}
\label{Tab:all_results}
\scalebox{1}{
\begin{tabular}{ccccccccccccccc}
\toprule
\multirow{2}{*}{Model} &\multirow{2}{*}{Method}&\multirow{2}{*}{NRU}&\multicolumn{3}{c}{\textbf{Coat}}&\multicolumn{3}{c}{\textbf{Yahoo! R3}} &\multicolumn{3}{c}{\textbf{KuaiRec}}\\\cmidrule{4-12}
 & & & MSE& MAE & nDCG@5 & MSE & MAE & nDCG@5 & MSE & MAE & nDCG@5\\ \midrule 
 \multirow{9}{*}{MF}
&\multicolumn{1}{c}{HEI} & ${\checkmark}$ &1.3614	&0.8927	&0.7785	&2.3874&1.2127&0.8013 &2.3227 & 1.3705 & 0.3448\\
&\multicolumn{1}{c}{Naive} & ${\checkmark}$& 1.2851& 0.8704 & 0.7850 & 2.3887& 1.2131 & 0.8015 &2.3817 &	1.3935 &0.3326 \\
&\multicolumn{1}{c}{SNIPS-UI} & ${\checkmark}$& 1.2342 & 0.8564 & 0.7846 & 1.9772 & 1.0519 & 0.7774 &1.2084 & 0.8663 & 0.3255\\
&\multicolumn{1}{c}{DR-UI} &${\checkmark}$ & 1.2330 & 0.8486 & 0.7704 & 2.5912 & 1.2543	& 0.7813 & 2.0233 & 1.0457 & 0.3466 \\
&\multicolumn{1}{c}{AT} & ${\checkmark}$& 1.1603	& 0.8294	& {0.7912} & 1.8389 &	0.9613	& 0.7973 &1.2274 & 0.8752 
& 0.3622 \\ 
&\multicolumn{1}{c}{CVIB} & ${\checkmark}$&1.2000 &0.9025 & 0.7232  & 1.1270 & 0.8537 & 0.7321 & 1.1940 &  0.9733 & 0.3327 \\ 
&\multicolumn{1}{c}{ESCM2} & $\checkmark$ &1.1368 &
0.8173 &
0.7976 & 1.1675 &
0.8154 &
\underline{\textbf{0.8157}} & 2.0414 & 1.2712 & 0.5403 \\
&\multicolumn{1}{c}{CounterCLR} & ${\checkmark}$
& \textbf{1.0956} &
\underline{\textbf{0.8008}} &
\underline{\textbf{0.8002}} &
\underline{\textbf{1.1137}} & {\textbf{0.8117}} & {0.8049}  &\underline{\textbf{1.1586}}  &	\underline{\textbf{0.8523}} &	\textbf{0.5893} \\ \cmidrule{2-12}
&\multicolumn{1}{c}{RDC} & $\times$ & \underline{1.0946}&
{0.8097} & {0.7930} & 1.3260 &0.8521 & 0.8040 & 1.2733 & 0.9297
& {0.5941}\\ 
&\multicolumn{1}{c}{SNIPS-NB}& $\times$& 1.1770 & 0.8367 & 0.7720 &	1.3314  &	0.8524 & 0.8039 & 1.3599 & 0.9692 & 0.4920
 \\
&\multicolumn{1}{c}{DR-NB} & $\times$& 1.2254 & 0.8673 & 0.7617 & 1.1875 & 0.8278 & {0.8058} & 1.4308 & 0.9872 & 0.5484 \\
&\multicolumn{1}{c}{SDR-NB} & $\times$ & 1.1923	&0.8645 &	0.7749 & 1.1599 &	\underline{0.8103}	&{0.8125} & 1.2793 & 0.9053 & \underline{0.6102}\\
% &\multicolumn{1}{c}{MTL}  \\
\midrule
 \multirow{9}{*}{NCF}
&\multicolumn{1}{c}{HEI} & ${\checkmark}$ &1.3557	&0.9388	&0.7321	&2.1483	&1.2070	&0.7995 & 0.4239 & 0.5009 & 0.3576\\
&\multicolumn{1}{c}{Naive} &${\checkmark}$ & 1.4343 &	0.9583 &	0.7320 & 2.1967 &	1.2521 &	0.7963 &0.4268
&0.5279
&0.4252 \\
&\multicolumn{1}{c}{SNIPS-UI}& ${\checkmark}$&1.3052 &0.9228 &0.7341	& 2.3317 & 1.2412 &0.7930 & 0.5735
&0.5978
&0.3439
 \\
&\multicolumn{1}{c}{DR-UI} &${\checkmark}$ & 
1.3998 &0.9453 &0.7347&2.7590 &1.3169	& 0.7769 & 0.3517 & 0.4243 & 0.6093 \\
&\multicolumn{1}{c}{AT}&${\checkmark}$ & 1.2641 & 0.9102 & 0.7402	  	& 2.2167 &	 1.2346 &	0.7961 & 0.5030
&0.5958
&0.3183  \\ 
&\multicolumn{1}{c}{CVIB} &${\checkmark}$&  1.2197 & 0.8893 &0.7265	&1.2505
& 0.9795
& 0.7895 & 0.3978 &0.4774 & 0.5298 \\ 
&\multicolumn{1}{c}{ESCM2} & ${\checkmark}$  & 1.2505 &
0.9105 &
0.7410 & 2.1919 &
1.2279 &
0.7999 & 0.4780 & 0.4973 & 0.5916\\
&\multicolumn{1}{c}{CounterCLR}&${\checkmark}$ & \underline{\textbf{1.1743}} & \textbf{0.8733} & \underline{\textbf{0.7421}} &	{\textbf{1.2203}}	& \underline{\textbf{0.8180}} &	\underline{\textbf{0.8040}}  & \textbf{0.3425} & 
\textbf{0.4129} &
\textbf{0.6198}
 \\  \cmidrule{2-12}
&\multicolumn{1}{c}{RDC} & $\times$& 
1.2857 & 0.9147 & {0.7415}	& 2.0037
& 1.2411
& 0.8006 & 0.3590 &0.4389
& 0.6193 \\
% &\rowcolor{lightgray} 
&\multicolumn{1}{c}{SNIPS-NB}&$\times$ &
1.2353	& 0.8788 &	0.7376	& {1.0838} & 0.8311 &  0.7948 & 0.3949 & 0.4820 & 0.6050\\
&\multicolumn{1}{c}{DR-NB}& $\times$& 
{1.1760} & \underline{0.8593}	&0.7362	& {1.3213} &0.8815	& 0.7913 &  \underline{0.3302} &\underline{0.3874} & \underline{0.6257}\\
&\multicolumn{1}{c}{SDR-NB} & $\times$  &1.2330 &0.8824 	&0.7367	&\underline{1.0821} &0.8195 &0.8027 & 0.3865 & 0.4323 & 0.6180\\
% &\multicolumn{1}{c}{MTL} & $\times$ \\\\
\bottomrule
\end{tabular}
\vspace{-6pt}}
%\end{spacing}
\end{table*}

% \vspace{-5mm}

\subsection{Overall Loss}
% \label{subsec: ol}
Taking the above two components together,
% causal rating model and contrastive learning objective together, 
{
we define the overall loss $\mathcal{L}$ as
\begin{equation}\label{eq: overall}
    \mathcal{L} = \mathcal{L}_{cau} + \beta \mathcal{L}_{con} = \mathcal{L}_{base} + \alpha \mathcal{L}_{pro} + \beta \mathcal{L}_{con},
\end{equation}
where $\beta$ is a hyper-parameter to control the contribution of the contrastive learning objective.

\section{Experiment}
\label{sec:exp}

% We compare the proposed CounterCLR with several baseline approaches, and conduct parameter sensitivity analysis to study how the parameters affect the performance of CounterCLR. 
% Three real-world datasets are used in our experiments.
% are the Coat dataset and Yahoo dataset.
In this section, we empirically validate the performance of CounterCLR by answering the research questions (RQs):
\textbf{RQ1.} Does the proposed CounterCLR achieve the state-of-the-art capability in mitigating selection bias in rating prediction task?
\textbf{RQ2.} Does the proposed CounterCLR outperform the baselines with varying data sparsity level?

% \begin{itemize}[leftmargin=*]
% \item \textit{(RQ1)} Does the proposed CounterCLR achieve the state-of-the-art performance in handling selection bias in recommendation task?
% % \item \textit{(RQ2)} What are the potential sources of gain in debiasing for the proposed CounterCLR?
% \item \textit{(RQ2)} How does the proposed method perform against the baselines under data sparsity? 
% \end{itemize}

\subsection{Experiment Setup}
\label{subsec: Experiment Setup}

% {\color{blue}We first introduce the datasets and comparison methods used in our experiments. The implementation details can be found in Appendix.}

% \noindent$\bullet$\hspace{3mm}\textit{Coat Shopping Dataset\footnote{{https://www.cs.cornell.edu/~schnabts/mnar/}}.} 

% \begin{table}[h] \footnotesize
% \caption{Statistics of the datasets. $\#R$ and $OR$ represent the number of ratings and  the observed ratio, respectively.}
% \label{Tab:datasets}
% \begin{spacing}{1.2}
% \begin{tabular}{ccccccc}\hline \hline
%       & \multicolumn{2}{c}{Coat} & \multicolumn{2}{c}{Yahoo} & \multicolumn{2}{c}{KuaiRec} \\ \cline{2-7}
%       & $\#R$    & $OR$   & $\#R$    & $OR$   & $\#R$      & $OR$    \\ \hline
% Train & 6,960        & 8\%       & 311,704      & 2\%        & 12,530,806     & 16.3\%     \\
% Test  & 4,640        & 5.3\%     & 54,000       & 0.4\%      & 4,676,570      & 99.6\%  \\  \hline \hline
% \end{tabular}
% \end{spacing}
% \end{table}

% \subsubsection{Datasets}
% \textbf{Datasets.} 
% To evaluate the performance in handling selection bias and data sparsity issues, 
We consider three real world datasets: \textbf{Coat}\footnote{https://www.cs.cornell.edu/\textasciitilde schnabts/mnar/}}, \textbf{Yahoo! R3}\footnote{{http://webscope.sandbox.yahoo.com/}}, and \textbf{KuaiRec}\footnote{{https://kuairec.com/}}.
All datasets contain biased training data acquired through traditional data collection process and unbiased test data acquired through randomized controlled trials (\textbf{Coat}, \textbf{Yahoo! R3}), and full exposure of test items (\textbf{KuaiRec}).
For methods requiring unbiased data for training or propensity estimation, we randomly split out $5\%$ test data and use Naive Bayes propensity estimator (NB) and User-Item propensity
estimator (UI). We adopt three widely-used metrics, namely MSE, MAE, and NDCG@5, for performance evaluation and report the mean results over five runs.

% \vspace{-3mm}

We choose representative debiasing methods as baselines, including {HEI} \cite{steck2010training}, {Naive} \cite{koren2009matrix},  {DR} \cite{saito2019doubly},  {AT} \cite{saito2020asymmetric}, {CVIB} \cite{wang2020information},
{ESCM2} \cite{wang2022escm2}, {RDC} \cite{liu2022rating}, and
{SDR} \cite{li2023stabledr}.
All methods are taking Matrix Factorization (MF) model and Neural Collaborative Filtering (NCF) model as base model on Pytorch with Adam as the optimizer. We use cross validation to select the learning rate in $\{5e-3, \cdots ,1e-1\}$,  weight decay in $\{1e-7, \cdots, 1e-1\}$, and batch size in $\{128, \cdots ,4096\}$. For the proposed CounterCLR, we also use cross-validation to select $K$ among $\{5,10,20,30,40\}$, $\alpha$ among $\{0.1, 1, 5, 10, 50, 100, 200\}$, and $\beta$ among $\{0.01, 0.1, 1, 5, 10\}$.
We set the momentum number $m$ as 0.999, the  hyper-parameter $\tau$ as 1, and temperature hyper-parameter $t$ as 0.07 by following \cite{liu2022rating} and \cite{he2020momentum}.

% In addition, for AT, we follow \cite{saito2020asymmetric}
% to take SNIPS method with UI estimator as the backbones of the two pre-trained models and target model for a fair comparison. For ESCM2, the choice of inverse propensity score and doubly robust as counterfactual risk minimizer demonstrates similar performance, and we use the former to show the performance.

% \begin{table}[h]\footnotesize
% \centering
% \setlength{\tabcolsep}{3pt}
% \begin{spacing}{1.2}
% \caption{{Rating prediction accuracy and recommendation
% quality(with mean and standard deviation) of the ablation studies over Coat and Yahoo}. The best MSE, MAE, and nDCG@5 are bold.}\label{Tab:ablation}
% \begin{tabular}{ccccccc}
% \hline \hline
% \multirow{2}{*}{Model} &\multicolumn{3}{c}{\emph{Coat}}&\multicolumn{3}{c}{\emph{Yahoo}}\\\cline{2-7}
% &  MSE& MAE & nDCG@5 & MSE & MAE & nDCG@5 \\ \hline 
% MF-CounterCLR & \textbf{1.0956}
% &\textbf{0.8008}
% &\textbf{0.8002} & \textbf{1.1137} & \textbf{0.8117} 	& \textbf{0.8049}\\
% w/o $\mathcal{L}_{pro}$ & 1.1199
% &0.8086
% &{0.7947} &	1.3170	& 0.8460 &	0.8038 \\
% w/o $\mathcal{L}_{con}$ &1.1012
% &0.8153
% &0.7921 &1.1237 &	0.8125 & 0.8046\\
% \hline
% NCF-CounterCLR & \textbf{1.1743} & \textbf{0.8733} & \textbf{0.7421} &	\textbf{1.2203}	&  \textbf{0.8180} &	\textbf{0.8040}\\
% w/o $\mathcal{L}_{pro}$ & 1.2635 &	0.9052	& 0.7384 & 2.1897 & 1.2330 & 0.8026 \\
% w/o $\mathcal{L}_{con}$ & 1.1755 &	0.8748 &	0.7419 & 1.3312 & 0.8528 & 0.8039\\
% \hline \hline
% \end{tabular}
% \end{spacing}
% % \vspace*{-0.1in}
% % \vspace*{-0.1in}
% \end{table}

% \vspace{-3mm}

\subsection{Experiments Results (RQ1)}
% \label{subsec: results}

% Table \ref{Tab:all_results} reports the experimental results of recommendation performance for all methods on the Coat dataset and the Yahoo dataset from the rating prediction accuracy and recommendation quality, and Table \ref{Tab:moive_results} shows the results on the Movie. 
{
% We follow \cite{schnabel2016recommendations} to measure the rating prediction accuracy by Mean Absolute Error (MAE) and Mean Squared Error (MSE), and measure the recommendation quality by Normalised Discounted Cumulative Gain at 5 (nDCG@5). 
% We report the MAE, MSE, and nDCG@5 results for all methods on the Coat dataset and the Yahoo dataset (Table \ref{Tab:all_results}), and the KuaiRec dataset (Table \ref{Tab:moive_results}).
% We also report the mean and standard deviation results over five runs for all methods in Tables \ref{Tab:all_results} and \ref{Tab:moive_results}.
% Note that smaller MAE and MSE imply better rating prediction accuracy, while a larger nDCG@5 suggests better recommendation quality.

The debiasing performance is shown in Table \ref{Tab:all_results}. First, the proposed CounterCLR outperforms the baseline methods both with unbiased data and without unbiased data on all three datasets, which suggests that the integration of the CauNet and the contrastive learning objective does facilitate reducing the influence of the selection bias. Second, we note that SNIPS-NB and DR-NB methods perform better than SNIPS-UI and DR-UI methods, which is because that NB can provide a more accurate propensity score estimation since the additional 5\% unbiased MAR test data. It verifies that the propensity-based methods rely on an accurate propensity score estimator to achieve desirable performance. The proposed CounterCLR, on the other hand, does not requires additional unbiased MAR data, and thus is more robust and practical.
}

% \subsection{Ablation Study (RQ2)}
% \label{subsec: ablation}
% {
% We investigate the influence of $\mathcal{L}_{pro}$ and $\mathcal{L}_{con}$ on the recommendation performance over Coat and Yahoo datasets. 
% The corresponding experimental results are shown in Table~\ref{Tab:ablation}. 
% It is obvious that, each element, i.e., the propensity estimation head in CauNet and the contrastive learning objective, does have a positive effect on the recommendation quality. 
% In particular, we can observe that removing $\mathcal{L}_{pro}$ leads to the worst performance in most cases, which verifies that our CauNet does model the causality in rating prediction task to alleviate the selection bias.
% Moreover, the performance degradation of removing $\mathcal{L}_{con}$ also demonstrates that the contrastive learning objective can {help better learn the causality for rating prediction and} boost debiasing learning.
% }

\begin{figure}[h!]
\center
  \includegraphics[width=1\linewidth, height=0.35\textwidth]{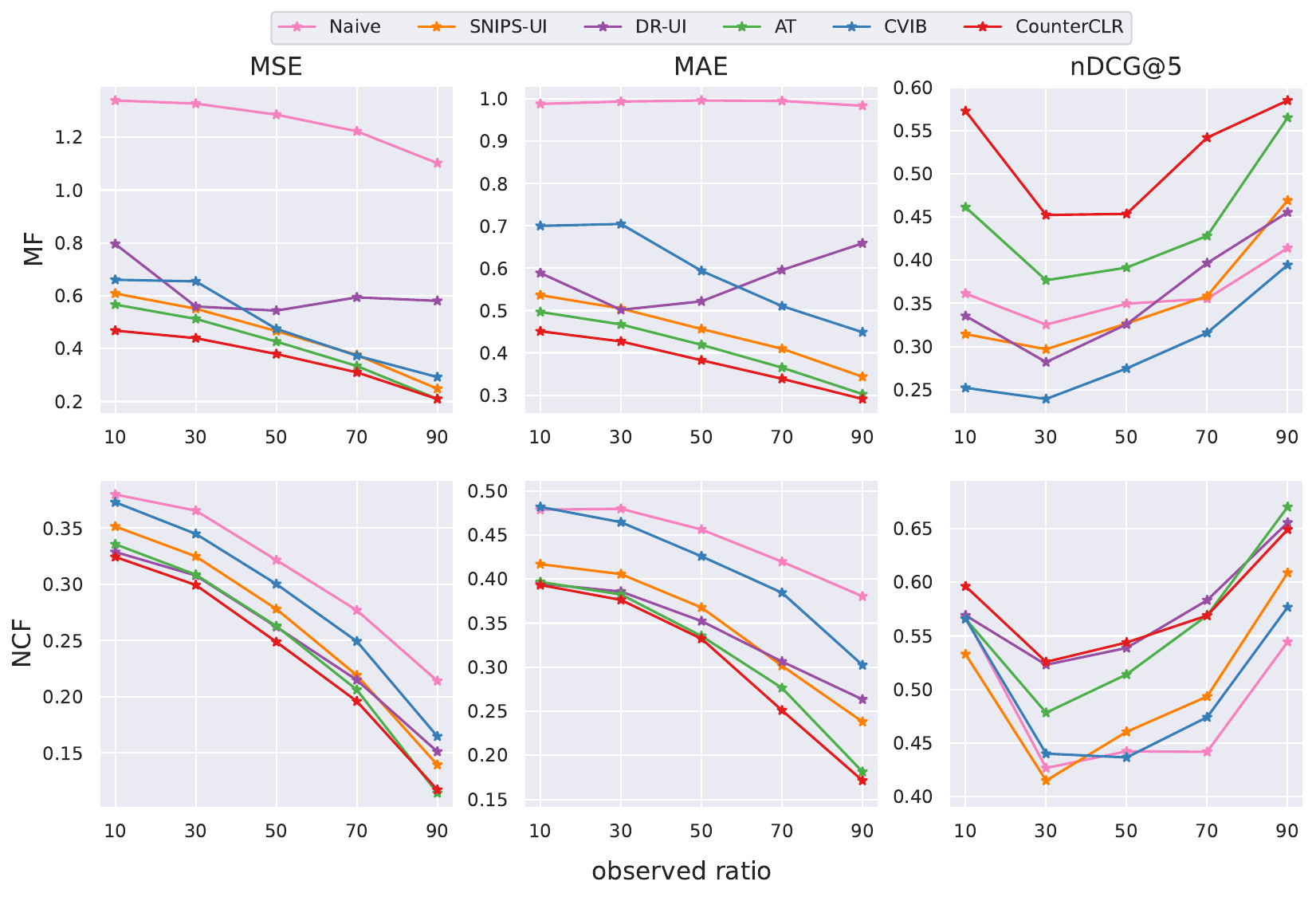}
  \vspace*{-6mm}
  \caption{Rating prediction accuracy and recommendation quality with varying observed ratio of the Small matrix in KuaiRec Dataset. }
 \label{fig:datasparsity}
 \vspace*{-3mm}
\end{figure}

%\vspace{-1.5mm}

\subsection{Influence of Data Sparsity (RQ2)}
\label{subsec: sparsity}
{
In order to evaluate the recommendation performance under data sparsity issue, we exploit the ``full observed'' \textit{small matrix} in \textbf{KuaiRec} to synthesize partially-observed data, with the observed ratio varying in $\{10\%, 30\%, 50\%, 70\%, 90\%\}$. Meanwhile, we follow the positive-oriented exposure strategy in \cite{gao2022kuairec} to simulate MNAR training data to model different selection bias level. % containing selection bias.
The unobserved parts in \textit{small matrix} are used as the test sets to evaluate the imputed missing data.
% First, we calculate the positive distribution on \textit{big matrix} according to the principle that the exposure probability of an item is proportional to the number of users liking it.
% Then, for each user in \textit{small matrix}, we follow the positive distribution to sample a certain number of items without replacement.
% In similar,  
% In this way, we obtain 5 MNAR training sets containing selection bias,
% %with the observed ratio equaling to $\{10\%, 30\%, 50\%, 70\%, 90\%\}$,
% and we take the unobserved parts corresponding to the 5 training sets in \textit{small matrix} as the test sets to evaluate the imputed missing data.

{We compare our CounterCLR with the baselines without unbiased MAR data for a fair comparison.}
% Figure~\ref{fig:datasparsity} reports the mean of MSE, MAE and nDCG@5 over five runs 
% results (MSE, MAE and nDCG@5) 
% in the five synthetic datasets.
From Figure~\ref{fig:datasparsity}, we first can observe that our CounterCLR stably outperforms the baselines in both rating prediction accuracy and recommendation quality. This shows the superior generalization ability of CounterCLR under data sparsity issues. Besides, with the observed ratios increasing, the curves of our CounterCLR and all the baselines show a downward trend in terms of MSE and MAE, and an upward trend in terms of nDCG@5.
This is because the higher observed ratio, the more training data will be accessed to train the recommender systems for providing more accurate recommendations. 
}

%\vspace{-3pt}
\section{Conclusion}
\label{sec:con}
%\vspace{-3pt}
In this work, we propose a novel causality-based contrastive learning framework CounterCLR for debiased rating prediction, which consists of a causal network CauNet and a contrastive learning objective. 
The proposed method CounterCLR can effectively address the selection bias and data sparsity issues simultaneously without introducing separate imputation and propensity estimators, or unbiased MAR data.
% , to deal with the selection bias and data sparsity issues.
Extensive experiments on real-world datasets show that CounterCLR outperforms other state-of-the-art rating prediction methods. 
% The development of a new, state-of-the-art technique for rating prediction can have transformative impacts on  learning better user preference and providing more accurate recommendation in recommender systems.

\newpage
\bibliographystyle{IEEEtran}
\bibliography{sample-base}

\end{document}